\documentclass[%
showpacs,
preprint,
amsmath,amssymb,
floatfix,
]{revtex4-1}
\usepackage{graphicx}
\usepackage{bm}
\usepackage[mathlines]{lineno}
\begin{document}
\title{Decay properties of a class of doubly charged Higgs bosons}
\author{M. D. Tonasse\footnote{Email: tonasse@registro.unesp.br.}}\thanks{On leave from Campus Experimental de Registro, Universidade Estadual Paulista, Rua Nelson Brihi Badur 430, 11900-000 Registro, SP, Brazil.}
\affiliation{Instituto de F\'\i sica Te\'orica, Universidade Estadual Paulista, Rua Dr. Bento Teobaldo Ferraz 271, 01140-070 S\~ao Paulo, SP, Brazil}
\date{\today}
\begin{abstract}
We study the leptonic decays of a doubly charged Higgs bosons class which is predicted by a model based on the
SU(3)$_C$$\otimes$SU(2)$_L$$\otimes$U(1)$_N$ electroweak gauge symmetry. In contrast to other models, decays into $\tau^\pm\tau^\pm$ are largely dominant (99.5 \% or more). Coupling of these scalars to two standard charged gauge bosons are either zero or very suppressed. Couplings to two different flavor of charged leptons do not occur. Some coupling features imposed by symmetry and representation content lead to simple relationships between decay rates and doubly charged Higgs masses. Some of the parameters depend only on the decay widths and on the charged lepton masses. In order to clarify the relevance of our results, some aspects of this model are compared with the Higgs triplet model.
\end{abstract}
\maketitle
{\it Introduction.}$-$ After a long search, the signs observed in ATLAS and CMS detectors at the LHC finally indicated the presence of a Higgs boson with mass around 125 GeV \cite {CM12}. This confirms the scheme of spontaneous symmetry breaking as a viable mechanism for generating masses. However, beyond the standard model (SM) there are a large number of electroweak models that predict the existence of neutral Higgs bosons. Therefore, although we cannot know yet which of these models belongs the discovered scalar, the $5.0$ standard deviations level of significance from the combined analysis of the measures, practically leaves no doubt about existence of a Higgs boson. In addition, the Collaborations D$0$ and CDF at Tevatron also just announced results compatible with the found by ATLAS and CMS \cite{CD12}. \par 
The spontaneous symmetry breaking plays an essential role in renormalizable gauge theories. In the SM, only one SU(2) complex scalar isodoublet is introduced. It is enough to break the symmetry and generate the weak boson masses via the Higgs mechanism and also the fermion ones (except for neutrinos) through appropriate Yukawa couplings. As a result, one unique neutral massive scalar boson remains after the spontaneous symmetry breaking. However, the standard Higgs boson is very elusive. It is electrically neutral, weakly interacting, and has many possible decay channels. Therefore, from the experimental point of view it would be interesting to find a more friendly scalar field which, although not being the standard Higgs, would provide a major advance in understanding the symmetry breaking and mass generation mechanisms. In fact, since it has been well established that the SM is only an approximate theory at low energy, much attention has also been given to scalars predicted by extended models. In many of them, extra scalar multiplets must be introduced to generate the masses to new particles. This can require the presence of physical charged scalars with one and/or two units of electric charge in the theory. This is the case of popular extensions of the SM, such as the left-right symmetric \cite{RI82}, little Higgs \cite{AC01}, the Higgs triplet (HTM) \cite{GV90}, and some versions of 3-3-1 model \cite{PT93, FH93, PP92, FR92}. \par
Doubly charged Higgs bosons (DCHBs) are interesting particles to be investigated in accelerators. They would be characterized by decay channels into two leptons with the same electric charge, which may give a particularly distinctive signal. Moreover, these processes violate the leptonic number, and through a convenient selection of cuts and decay modes, they can be made practically free from being masked by background effects \cite{AC10,HM07}. In addition, they could manifest themselves near the Fermi scale and could be easily detected at the LHC. The DCHBs of the left-right symmetric model and HTM were studied at the LHC. The CMS Collaboration presented bounds for HTM. Concerning to the searches in the channels $e^\pm e^\pm$, $\mu^\pm\mu^\pm$ and $e^\pm\mu^\pm$, it was obtained the bound $m_{H^{\pm\pm}} > 300$ GeV. For the channels $\mu^\pm\tau^\pm$ and $e^\pm\mu^\pm$ were found the limits $m_{H^{\pm\pm}} > 266$ GeV and $m_{H^{\pm\pm}} > 254$ GeV, respectively \cite{CM11b}. Left-right DCHBs were investigated by the ATLAS Collaboration through the $\mu^\pm\mu^\pm$ channel. The supposition of branching ratio ($BR$) of $100 \%$ resulted in the limits of $m_{H_L^{++}} > 375$ GeV and $m_{H_R^{++}} > 295$ GeV. For $BR = 33 \%$ the limits obtained were $m_{H_L^{++}} > 268$ GeV and $m_{H_R^{++}} > 210$ GeV \cite{AT11b}. \par
The aim of this work is to study the decays of some DCHBs predicted by the minimal 3-3-1 model into same electric charge leptons. The model has some features that make it unique with respect to the properties of the DCHB sector. The rates of the decay $H^{\pm\pm} \to W^\pm W^\pm$ are either zero or very suppressed. There is also no decay $H^{\pm\pm} \to \ell^\pm \ell^{\prime\pm}$ with the leptons $\ell \neq \ell^\prime$. The decay $H^{\pm\pm} \to \tau\tau$ is strongly favored. The minimal 3-3-1 model predicts three DCHBs (and its antiparticles). Two of them have charged lepton couplings. In this work we will show that there are simple relations between the DCHB masses and the respective decay rates. \par
{\it Relevant Features of the Model.}$-$ The 3-3-1 model belongs to a class of electroweak models based on SU(3)$_C$$\otimes$SU(3)$_L$$\otimes$U(1)$_N$ gauge symmetry. Their main virtue is that anomalies do not cancel independently in each generation, as in the SM. The anomaly cancellation is a model property only when the three generations are taken into account together. The cancellation mechanism requires the family number to be an integer multiple of the color number. Then, remembering that the QCD asymptotic freedom property dictates that the color number is less than five, it can be concluded that the 3-3-1 model predicts the existence of three and only three fermion families in nature. In addition to the 3-3-1 model this result can be obtained only in a hexadimensional model by introducing a condition on the global anomalies cancellation \cite{DP01}. Moreover, according to 3-3-1 models the Weinberg angle has the constraint $\sin^ \theta_W < 1/4$ in the minimal version \cite{PP92}. Therefore, when $\sin^2\theta_W$ evolves to high values it is shown that the model loses its perturbative character in a scale about 8 TeV \cite{DI05}. Note that in other models the energy scale may be increased practically indefinitely. In the minimal 3-3-1 model, it is not interesting, although not forbidden. Then, the 3-3-1 model is one of the most attractive extensions of the SM and is phenomenologically well motivated to be probed in the LHC and other colliders. \par
The formalism used here is very simple. It consists basically of calculations of eigenstates and DCHB decay widths. Our results have, in some sense, a similarity with an HTM forecast, although in this case the result is for the neutrino sector. In HTM, the charged lepton masses are generated through an SU(2) scalar doublet as in the SM, while an extra SU(2) scalar triplet generates neutrino masses at tree level when its neutral component develops a vacuum expectation value (VEV). The same Yukawa constants which parametrize the neutrino mass matrix are present in DCHB couplings with charged leptons. Therefore, it can be inferred that leptonic $BR$ measurements of the DCHBs induce bounds on (but not unequivocally determine) neutrino masses in the HTM context \cite{HM07,KR08,GS08}. In fact, according to HTM, it is possible to determine some neutrino sector parameters from the DCHB decays into charged leptons, but only through statistical analysis \cite{GS08} or by assigning values to a free parameters set \cite{KR08}. On the other hand, the minimal 3-3-1 model provides simple relationships between the DCHB masses and the correspondent decays widths, as we will show later in this work. So long as it is possible to eliminate possible experimental difficulties, these results can be used to determine the mass values and other DCHB free parameters unambiguously. According to the best of our knowledge, no other electroweak model is capable to provide a result of this kind, at least with so few experimental data. As already mentioned, the DCHB decays into $W^\pm W^\pm$, which can be dominating in the HTM, are very unlike in this context. There is also no decays into leptons of different flavors. As soon, as there is no reason to believe that the new boson masses are very disparate, we will work with purely leptonic decays into same flavor charged leptons. \par
To study DCHB leptonic decays we started with the Yukawa couplings, which are given in the minimal 3-3-1 model by
\begin{equation}
-{\cal L}_1 = \frac{1}{2}\sum_{a, b}G_{ab}\overline{\left(\psi_{iaL}\right)^c}\psi_{jbL}S^{ij} + {\mbox{H. c.}},
\label{L}\end{equation}
where $G_{ab}$ ($a$ and $b$ are generation indexes) are Yukawa constants. The leptons transform as
\begin{equation}
\psi_{aL} = \left(\begin{array}{c}\nu^\prime_a \\ \ell^\prime_a \\
\ell_a^{\prime c}\end{array}\right)_L \sim \left({\bm 3}, 0\right),
\label{psi}\end{equation}
under SU(3)$_L$. The introduction of SU(3)$_L$ right-handed leptonic singlets is not necessary, since $\left(\ell^c_a\right)_L = \left(\ell_{aR}\right)^c$. The interaction eigenstates in Eq. (\ref{psi}) can be rotated into their respective physical eigenstates by rotation matrices $U^{L(R)}$ suitably selected, such that
\begin{equation}
\ell^\prime_{aL(R)} = U^{L(R)}_{ab}\ell_{bL(R)}, \qquad
\nu^\prime_{aL(R)} = U^{L(R)}_{ab}\nu_{bL(R)}.
\label{giro}\end{equation}
The scalar sextet $S$ in Lagrangian (\ref{L}) is
\begin{equation}
S = \left(\begin{array}{ccc}\sigma^0_1 & s_2^+ & s_1^- \\
s_2^+ & S_1^{++} & \sigma^0_2 \\
s_1^- & \sigma^0_2 & S_2^{--}\end{array}\right) \sim
\left({\bm 6}, 0\right).
\label{S}\end{equation}
Indeed, the Higgs fields defined in Eq. (\ref{S}) are not sufficient to break the symmetry and generate the correct set of masses in the model. If we wish to maintain the minimum leptonic content, namely the one provided in Eq. (\ref{psi}), at least three other SU(3) scalar multiplets,
\begin{widetext}
\begin{equation}
\eta = \left(\begin{array}{c}\eta^0 \\ \eta^-_1 \\
\eta^+_2\end{array}\right) \sim \left({\bf 3}, 0\right), \qquad \rho = \left(\begin{array}{c}\rho^+ \\ \rho^0 \\ \rho^{++}\end{array}\right) \sim \left({\bf 3}, 1\right), \qquad \chi = \left(\begin{array}{c}\chi^-\\ 
\chi^{--} \\ \chi^0\end{array}\right) \sim \left({\bf 3}, -1\right)
\label{tri}\end{equation}
\end{widetext}
should be introduced \cite{FR92, FH93}. Then the scalar triplet $\eta$ would couple the leptons as $\epsilon^{ijk}\overline{\left(\psi^c_{ia}\right)_R}\psi_{jbL}\eta_k$.
But since this field has no essential role in the leptonic sector, we can eliminate these couplings by imposing the set of discrete symmetries
\begin{equation}
\eta \to -\eta, \qquad \rho \to -\rho,
\label{s}\end{equation}
and the other fields transforming trivially. \par
The most general renormalizable Higgs potential is
\begin{widetext}
\begin{eqnarray}
V\left(\eta, \rho, \chi, S\right) & = & \mu_1^2\eta^\dagger\eta + \mu_2^2\rho^\dagger\rho + \mu_3^2\chi^\dagger\chi +
\mu_4^2{\mbox{tr}}\left(S^\dagger S\right) + \alpha_1\left(\eta^\dagger\eta\right)^2 + \alpha_2\left(\rho^\dagger\rho\right)^2 + \alpha_3\left(\chi^\dagger\chi\right)^2 + \cr
&& + \alpha_4{\mbox{tr}}\left[\left(S^\dagger S\right)^2\right] + \eta^\dagger\eta\left(\alpha_5\rho^\dagger\rho +
\alpha_6\chi^\dagger\chi\right) + \alpha_7\rho^\dagger\rho\chi^\dagger\chi + \alpha_8\eta^\dagger\rho\rho^\dagger\eta + \cr
&& + \alpha_9\eta^\dagger\chi\chi^\dagger\eta + \alpha_{10}\rho^\dagger\chi\chi^\dagger\rho + \left[\alpha_{11}\eta^\dagger\eta + \alpha_{12}\rho^\dagger\rho + \alpha_{13}\chi^\dagger\chi + \alpha_{14}{\mbox{tr}}\left(S^\dagger S\right)\right]\times \cr
&& \times{\mbox{tr}}\left(S^\dagger S\right) + \left[\left(\alpha_{15}\eta_i\eta_j + \alpha_{16}\rho_i^\dagger\rho_j\right)\left(SS\right)^{ij} + \left(\alpha_{17}\eta\eta + \alpha_{18}\rho^\dagger\rho\right){\rm tr}\left(S^\dagger S\right) + \right. \cr
&& \left. + {\rm H. c.}\right] + \alpha_{19}\eta^\dagger S^\dagger S\eta + \alpha_{20}\rho^\dagger S^\dagger S^{\rm T}\rho + \alpha_{21}\chi^\dagger SS^\dagger\chi + \alpha_{22}{\rm tr}\left(SS^\dagger SS^\dagger\right)+ \cr
&& + \left[f_1\varepsilon^{ijk}\eta_i\rho_j\chi_k + f_2\eta^\dagger_i\eta_jS^{ij} + f_3{\rm tr}\left(SSS\right) + {\rm H. c.}\right]
\label{pot}\end{eqnarray}
\end{widetext}
In the definition (\ref{pot}) the constants $\mu_i$ $\left(i = 1, 2, 3, 4 \right)$ and $f_j$ $\left(j = 1, 2, 3 \right)$ have mass dimension, while $\alpha_k$ $\left(k = 1, \ldots, 2 \right)$ are dimensionless. The presence of the trilinear term proportional to $f_1$ in the potential (\ref{pot}) is the reason for the particular transformation of the scalar $\rho$ in (\ref{s}). In some cases, if we remove this term from the scalar potential, the model will have a neutral massless scalar boson \cite{TO96}, which would be difficult to explain from the point of view of experimental results. The triplet $\chi$ in Eqs. (\ref{tri}) is supposedly heavy and governs the spontaneous symmetry breaking of SU(3)$_L$$\otimes$U(1)$_N$ to SU(2)$_L$$\otimes$U(1)$_Y$ of the SM, while $S$, $\eta$ and $\rho$ are responsible for breaking SU(2)$_L$$\otimes$U(1)$_Y$ $\to$ U(1)$_{\rm em}$ of the electromagnetism. The neutral scalars can develop their VEVs $\langle\sigma^0_1\rangle = v_1$, $\langle\sigma^0_2\rangle = v_2$, $\langle\eta^0\rangle = v$, $\langle\rho^0\rangle = u$ and $\langle\chi^0\rangle = w$. Therefore, the symmetry breaking pattern leads us to expect
\begin{equation}
v_1, v_2, v, u \ll w.
\label{ap}\end{equation}
Henceforth we will express the VEVs through the ratios $a = v/w$, $b = u/w$ and $c = v_2/w$. The Higgs potential (\r {pot}) provides the mass matrix
\begin{widetext}
\begin{equation}
M_{++}^2 = \left(\begin{array}{cccc}
-3af_1w/2b & -3af_1w/2 & \Lambda bcw^2 & \Lambda bcw^2 \\
-3af_1w/2 & -3abf_1w/2 & \alpha_{21}cw^2 & \alpha_{21}cw^2 \\
\Lambda bcw^2 & \alpha_{20}cw^2 & \left(\Lambda b^2 - \alpha_{21}\right)w^2/2 & 0 \\
\Lambda bcw^2 & \alpha_{21}cw^2 & 0 & -\left(\Lambda b^2 - \alpha_{21}\right)w^2/2
\end{array}\right)
\label{matcc}\end{equation}
\end{widetext}
for DCHBs in the basis $\left(\rho^{++}, \chi^{++}, S_1^{++}, S_2^{++}\right)$, where $\Lambda = 2\alpha_{16} + \alpha_{20}$ and we have omitted terms proportional to the VEV $v_1$. It should be noted that the mass matrix (\ref{matcc}) supplies two doubly charged Goldstone bosons $\left(G^{\pm\pm}\right)$ which will give rise to the doubly charged gauge boson masses predicted by the model \cite{FH93, PP92, FR92}. \par
In order to find the approximate eigenvalues of the mass matrix (\ref{matcc}), we take the characteristic equation and expand the coefficients for $a \approx b \approx 0$ and $c \approx 0$ retaining only the leading term, i. e.,
\begin{equation}
x^3 + \frac{3af_1w}{2b}x^2 - \frac{\alpha_{21}^2w^2}{4}x -
\frac{3\alpha_{21}^2af_1w^5}{8b} = 0.
\label{char1}\end{equation}
All expansion made in this work are for $a \approx b \approx 0$. The factor $c$ is also small, but not necessarily of the order of $a$ and $b$. The Eq. (\ref{char1}) provides, after expanding the solutions,
\begin{subequations}\begin{equation}
m_{1cc} = 0, \qquad M_{2cc} \approx 0, \qquad m_{3cc}^2 \approx -M_{4cc}^2 \approx \frac{\alpha_{21}}{2}w^2.
\end{equation}
We assume the mass $m_{3cc}$ as approximately correct and $m_{1cc}$ is exact. $M_{4cc}^2$ is discarded because of the wrong sign. Then, to find better approximations to $m_{2cc}$ and $m_{4cc}$, we write a cubic equation in its general form. In following, we take into account $m_{3cc}$ and compare the cubic equation coefficients with those of the exact characteristic equation for the matrix (\ref{matcc}). To find compact solutions physically acceptable we expand $m_{2cc}^2$ for $a \approx b \approx 0$ and $c \approx 0$ up to order $1/w^3$ and $m_{4cc}^2$ up to order $w$. Thus we have
\begin{equation}
m_{2cc}^2 \approx -\frac{36a^2c^2f_1^2\left(3f_1a + \alpha_{21}bw\right)}{\alpha_{21}^2b^3w}, \qquad
m_{4cc}^2 \approx -\frac{3waf_1}{2b}.
\end{equation}\label{mass1}\end{subequations}
The eigenstates are obtained by taking the expansion leading term for each physical field, i. e.,
\begin{widetext}
\begin{subequations}\begin{eqnarray}
\rho^{\pm\pm} & \approx & -bG^{\pm\pm} + 2\sqrt{2}bc\left(H_1^{\pm\pm} + \frac{\Lambda}{\alpha_{21}}H_2^{\pm\pm}\right) + H_3^{\pm\pm}, \\
\chi^{\pm\pm} & \approx & G^{\pm\pm} + 2\sqrt{2}c\left(-H_1^{\pm\pm} + \frac{\Lambda}{\alpha_{21}}b^2H_2^{\pm\pm}\right) + bH_3^{\pm\pm}, \\
S_1^{\pm\pm} & \approx & 2cG^{\pm\pm} + \frac{1}{\sqrt 2}\left(H_1^{\pm\pm} + H_2^{\pm\pm}\right) - \frac{2\Lambda}{\alpha_{21}}bcH_3^{\pm\pm}, \\
S_2^{\pm\pm} & \approx & -2cG^{\pm\pm} + \frac{1}{\sqrt 2}\left(-H_1^{\pm\pm} + H_2^{\pm\pm}\right) - \frac{2\Lambda}{\alpha_{21}}bcH_3^{\pm\pm}.
\end{eqnarray}\end{subequations}
\end{widetext}
Therefore, the Yukawa Lagrangian (\ref{L}), in terms of physical states $H^{\pm\pm}_{2, 3}$, becomes
\begin{equation}
-{\cal L}_2 = \frac{1}{2v_2}\overline{L^+}\left(1 - \gamma_5\right)ML^-\left(\frac{1}{\sqrt 2}H_2^{++} - 2\frac{\Lambda}{\alpha_{21}}bcH_3^{++}\right) + {\rm H. c.}
\label{L2}\end{equation}
where $M = {\mbox{diag}}\left(\begin{array}{ccc} m_e & m_\mu & m_\tau\end{array}\right)$ and $L^\pm = \left(\begin{array}{ccc}e^\pm & \mu^\pm & \tau^\pm\end{array}\right)^{\tt T}$. Therefore, as can be seen from Lagrangian (\ref{L2}), the scalars $H_2^{\pm\pm}$ and $H_3^{\pm\pm}$ do not couple different flavor of charged lepton, since the DCHB Lagrangian was diagonalized along with the mass matrix (\ref{matcc}). \par
The covariant derivatives are given by
\begin{widetext}
\begin{subequations}\begin{eqnarray}
{\cal D}_\mu\varphi_i & = & \partial_\mu\varphi_i - ig\left(W^a_\mu\frac{\lambda^a}{2}\right)^j_i\varphi_j - ig^\prime
N_\varphi\varphi_iB_\mu, \\
{\cal D}_\mu S_{ij} & = & \partial_\mu S_{ij} - ig\left[\left(W^a_\mu\frac{\lambda^a}{2}\right)^k_iS_{kj} + \left(W^a_\mu\frac{\lambda^a}{2}\right)^k_jS_{ki}\right],
\end{eqnarray}\label{cov}\end{subequations}
\end{widetext}
where $\varphi_i = \eta, \rho, \chi$, $W_\mu$ and $B_\mu$ are the SU(3)$_L$ and U(1)$_N$ gauge field tensors, respectively, and $g$ and $g^\prime$ are coupling constants related the gauge groups SU(3)$_L$ and U(1)$_N$. Thus, the interactions of the Higgs with the gauge bosons are described by
\begin{equation}
{\cal L}_H = \left({\cal D}_\mu S_{ij}\right)^\dagger\left({\cal D}_\mu S_{ij}\right) + \sum_\varphi\left({\cal D}_\mu\varphi\right)^\dagger\left({\cal D}_\mu\varphi\right).
\label{LH}
\end{equation}
Therefore, we can see that the parameter $\rho$ provides no bound applicable to the mass scales of the model, since that the standard gauge boson masses which can be deduced from the Lagrangian (\ref{LH}) are
\begin{equation}
m_W = \frac{g}{2}v_W, \qquad m_Z^2 \approx \frac{g^2}{4}\frac{v_W^2 + 2\left(v_1^2 + 2v_2^2\right)}{1 - s_W^2},
\end{equation}
where for $m_Z^2$ we used the approximation (\ref{ap}). \par
An important result, which can be obtained from the Lagrangian (\ref{LH}), is that the trilinear coupling of the DCHB with two charged standard gauge bosons is $g^2\sigma_1^0W^\pm W^\pm S_1^{\mp\mp}$, before the spontaneous symmetry breaking. But, from the Yukawa Lagrangian (\ref{L}) we can see that $\sigma_1^0$ is the scalar responsible for the neutrino masses generation. So, we hope that this coupling is very suppressed. It can even be zero if $v_1$ vanishes. In this case the neutrino masses can be generated by higher order processes. This is why we are assuming $v_1 = 0$ throughout this work. It should be emphasized here that the 3-3-1 model has never been experimentally probed. Therefore, there are no constraints on its free parameters unless it is necessary to be consistent with results at low energies. However, there is a motivation to take $v_2 < v, u$, since here the role of $v_2$ is essentially to give mass to the charged leptons. \par
{\it Decay Widths} $-$ The decay widths $\Gamma_{i\ell}$ $\left(i = 2, 3\right)$ of $H^{\pm\pm}_i \to \ell^\pm\ell^\pm$ can be written as
\begin{equation}
\Gamma_{i\ell} = \frac{1}{8\pi}\left(\frac{\Theta_{i\ell}}{m_i}\right)^2\left(m_i^2 - 2m_\ell^2\right)\sqrt{m_i^2 - 4m_\ell^2},
\label{gam}\end{equation}
where $m_i$ are the DCHB masses and $\Theta_{i\ell}$ are the coupling strengths for $H_i^{\pm\pm}\ell^\mp\ell^\mp$ interaction which can be read from the Lagrangian (\ref{L2}). Therefore, from Eq. (\ref{gam}) we can write immediately
\begin{widetext}
\begin{subequations}\begin{eqnarray}
v_2^2 & \approx & \frac{1}{128\pi\Gamma_{2\mu}}\left(\frac{m_\mu}{m_\tau}\right)^2\sqrt{m_2^2 - 4m_\mu^2}\left(m_2^2 - 3m_\mu^2\right), \\
m_{i\pm}^2 & \approx & 2\frac{2\left(\kappa_im_\mu^2 - m_\tau^2\right)\pm \sqrt{4\left(\kappa_im_\mu^2 - m_\tau^2\right)^2 - 5\left(\kappa_i - 1\right)\left(\kappa_im_\mu^4 - m_\tau^4\right)}}{\kappa_i - 1},
\label{ms}\end{eqnarray}\label{res}\end{subequations}
\end{widetext}
where $\kappa_i = \left(\Gamma_{i\tau}m_\mu^2/\Gamma_{i\mu}m_\tau^2\right)$ $\left(i = 2, 3\right)$. In the derivation of  Eq. (\ref{ms}) we neglected cubic terms in $m_\mu$ and $m_\tau$. \par
{\it Results.}$-$ The DCHBs of the 3-3-1 model have some unusual features when compared with other models, which we summarize below:
\begin{enumerate}
\item The decays into two leptons of the same charge occur almost entirely in tau pairs $\left[BR\left(H_{2,3}^{\pm\pm} \to \tau^\pm\tau^\pm\right) > 99.5 \% \right]$.
\item On the other hand, $H_{2,3}^{\pm\pm} \to W^\pm W^\pm$ are disfavored because the coupling $H_{2,3}^{\pm\pm}W^\mp W^\mp$ is proportional to the VEV of the Higgs which gives mass to neutrinos.
\item The decays $H_{2,3}^{\pm\pm} \to \ell^\pm\ell^{\prime\pm}$, with $\ell \neq \ell^\prime$, do not occur because the Yukawa couplings are diagonalized together with the diagonalization of DCHB mass matrix.
\item There are simple and firmly established relationships between mass parameters and the respective DCHB decay widths. Eqs. (\ref{gam}) and (\ref{res}) have two remarkable feature:
\begin{enumerate}
\item The quantities $v_2$ and $m_{i\pm}$ contain no free parameters other than the decay widths. It should be noted that mixing angles, as those defined in Eqs. (\ref{giro}), do not appear, as would be the case in the HTM. Therefore, one possible measure of some of these widths would lead to an unambiguous determination of these quantities. In other
extensions of the SM only upper limits can be obtained at accelerators.
\item On the other hand, Eq. (\ref{gam}) gives $\Gamma_{i\ell} \approx \Gamma_{i\ell^\prime}m^2_{\ell}/m^2_{\ell^\prime}$. Therefore, these measurements are more difficult to be made in the context of the 3-3-1 model than the others, since there is a strong preference for the decay $H_i^{\pm\pm} \to \tau^\pm\tau^\pm$.
\end{enumerate}
\end{enumerate}
We can illustrate the results with a numerical example. For the charged leptons we take $m_e = 0.51$ MeV, $m_\mu = 105.66$ MeV, $m_\tau = 1776.82$ MeV, and we assume $\Gamma_{2\mu} = \Gamma_{3\mu} = 1.00$ GeV and $\Gamma_{2\tau} = \Gamma_{3\tau} = 282.79$ GeV as 3-3-1 model input values. Thus we get $m_{i+} = 300.00$ GeV, $v_2 = 0.091$ GeV. We have also $\Gamma_{2e} = \Gamma_{3e} = 2.33 \times 10^{-6}$ GeV.\par
This simple numerical example shows the difference in ability between the minimal 3-3-1 model and the other SM extensions, such as HTM, to predict the DCHB free parameter values. In this case, without taking into account the experimental difficulties, four DCHB decay widths (or $BR$s) need to be measured to calculate one of the VEVs, two DCHB masses and its decays widths into $e^\pm e^\pm$ pairs. \par
{\it Comments.}$-$ Let us discuss the possibility of detecting DCHBs 3-3-1 model, compared with the HTM, in the light of the standard detection techniques. So, to see if $H_{2,3}$ are interesting to be searched at the LHC we must examine the cross sections of its production processes (or, equivalently, the strength of the couplings involved) and the $BR$s. 
\begin{figure}[h]
\begin{center}
\includegraphics[width = 14 cm, height = 3 cm]{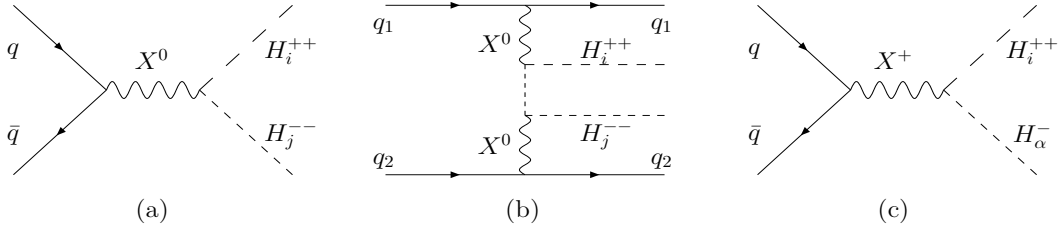}
\end{center}
\caption{\footnotesize\baselineskip = 12pt Diagrams for pair production of DCHBs through neutral bosons exchange (a) and bosons fusion (b) and single production through charged bosons exchange (c) in the minimal 3-3-1 model. In the intermediate states we have $X^0 = \gamma, Z, Z^\prime$ and neutral Higgs and $X^+ = W^+, V^+, H^+_\beta$. In this figure, $i, j = 2, 3$ and $\alpha, \beta = 1, 2, 3, 4$.}
\label{fig1}
\end{figure}
Obviously the cross sections depend also on the intermediate masses, but there are no constraints for the 3-3-1 mass scales. Therefore, the 3-3-1 boson masses can be compatible with the mass scales of the model in comparison. In Fig. 1 we present the main DCHB production processes of the minimal 3-3-1 model relevant for studies at the LHC. Therefore, we need of the singly charged Higgs boson physical eigenstates. From the Higgs potential (\ref{pot}) we have the singly charged sector mass matrix
\begin{subequations}
\begin{widetext}
\begin{equation}
M_+^2 = \left(\begin{array}{cccccc}
\beta & -f_1w & 0 & 0 & \alpha_{17}ac^2w & 0 \\
-f_1w & -af_1w/b & 0 & \alpha_{16}c^2w^2 & 0 & 0 \\
0 & 0 & \beta & 0 & 0 & \alpha_{17}acw^2 \\
0 & \alpha_{16}c^2w^2 & -bf_1w & \gamma & 0 & \alpha_{18}cw^2 \\
\alpha_{17}caw^2 & 0 & 0 & 0 & \alpha_{17}a^2w^2 & 0 \\
0 & 0 & \alpha_{17}acw^2 & \alpha_{18}cw^2 & 0 & \left(\alpha_{17}a^2 - \alpha_{18}\right)w^2
\end{array}\right)
\label{matc1}
\end{equation}
\end{widetext}
in the basis $\left(\eta^+_1, \eta^+_2, \rho^+, \chi^+, s_1^+,
s_2^+\right)$, where we have defined $\beta = \left(\alpha_{17}ac^2w - bf_1\right)w/a$ and $\gamma = -\left(\alpha_{18}c^2w + abf_1\right)w$.
The matrix (\ref{matc1}) gives us four singly charged Goldstone bosons $\left(G^\pm_{1,2}\right)$. To diagonalizes it, we take the first-order expansion terms of their elements except the $2 \times 2$ upper left corner submatrix. So, we get
\begin{widetext}
\begin{equation}
M_{+a}^2 = \left(\begin{array}{cccc}
\left(\alpha_{17}ac^2w - bf_1\right)w/a & -f_1w & 0 & 0 \\
-f_1w & -af_1w/b & 0 & 0 \\
0 & 0 & \alpha_{17}a^2w^2 & 0 \\
0 & 0 & 0 & \left(\alpha_{17}a^2 - \alpha_{18}\right)w^2
\end{array}\right),
\label{matc2}
\end{equation}
\end{widetext}
\end{subequations}
from which we have masses
\begin{subequations}\begin{eqnarray}
m_{c1}^2 \approx \alpha_{17}a^2w^2, && m_{c2}^2 \approx \left(\alpha_{17}a^2 - \alpha_{18}\right)w^2, \\
m_{c3}^2 \approx \alpha_{17}\frac{a^2c^2}{a^2 + b^2}w^2, && 
m_{c4}^2 \approx -\frac{a^2 + b^2}{ab}f_1w.
\end{eqnarray}\end{subequations}
With these eigenvalues and the exact mass matrix (\ref{matc1}) we find, also in first approximation, the eigenstates
\begin{widetext}
\begin{subequations}\begin{eqnarray}
\eta_1^\pm & \approx & -\frac{2\alpha_{15} + \alpha_{19}}{2}\frac{wc}{f_2}H_2^\pm - H_3^\pm, \\
\eta^\pm_2 & \approx & H^\pm_2 - \frac{2\alpha_{15} + \alpha_{19}}{2}\frac{wc}{f_2}H^\pm_3, \\
\rho^\pm & \approx & -\frac{\alpha_{21}\Lambda}{\Upsilon^-
}\frac{b}{c}G^\pm_1 - 32\left(\frac{\alpha_{18}}{\Upsilon^+}\right)^2\left(\frac{\Lambda\Omega}{\Upsilon^-}\frac{b}{c}\right)^3G_2^\pm - \frac{16\alpha_{21}}{3}\left(\frac{\alpha_{18}}{\Upsilon^-}\right)^4\Lambda^3\Omega\frac{wb^5}{f_3c^4}H_1^\pm + \cr
&& + \frac{4\alpha_{18}^2}{\Upsilon^+}\frac{a}{b}\left(H_2^\pm - H_3^\pm\right) - \frac{4\alpha_{21}\Lambda}{3}\left(\frac{\alpha_{18}}{\Upsilon^-}\right)^2\frac{wb^3}{f_3c^2}H_4^\pm, \\
\chi^\pm & \approx & -\frac{2\alpha_{18}\Lambda}{\Upsilon^-
}\frac{b}{c}G_1^\pm - 16\alpha_{18}^3\left(\frac{\Lambda^2}{\Upsilon^+}\right)^2\left(\frac{\Omega b}{\Upsilon^-c}\right)^3G^\pm_2 + \frac{8\alpha_{18}^3}{3}\frac{\Lambda\Omega}{\Upsilon^{-2}}\frac{b^3}{f_3c^2}H_1^\pm + \cr
&& + 2\frac{\alpha_{18}\Theta}{\Upsilon^+}\frac{a}{b}\left(H_2^\pm - H_3^\pm\right) + \frac{8\alpha_{18}^3\Lambda}{3\Upsilon^{-2}}\frac{b^3}{f_3c^2}H_4^\pm, \\
s_1^\pm & \approx & -G_1^\pm + 32\Lambda\left[\frac{\alpha_{18}}{\Upsilon^+}\left(\frac{\Omega}{\Upsilon^-}\right)^2\right]^2\left(\frac{b}{c}\right)^4G_2^\pm + \frac{4\alpha_{18}^2\Lambda}{3\Upsilon^-}\frac{wb^2}{f_3c}H_1^\pm - \frac{4\alpha_{18}^2\Lambda\Omega}{\Upsilon^-\Upsilon^+}\frac{a}{b}H_2^\pm + \cr
&& - \frac{\Theta^2a}{\Upsilon^+c}H_3^\pm + \frac{4\alpha_{21}}{3}\left(\frac{\lambda_{18}^2\Lambda}{\Upsilon^-}\right)^2\frac{wb^4}{f_3c^3}H_4^\pm \\
s_2^\pm & \approx & \frac{4\alpha_{18}^2\Lambda}{3\Upsilon^-
}\frac{wb^2}{f_3c}G_1^\pm + \frac{128}{3}\alpha_{18}\alpha_{21}\Omega\left(\frac{\Lambda}{\Upsilon^-\Upsilon^+}\right)^6\frac{wb^8}{f_3c^7}G_2^\pm + H_1^\pm + \cr
&& + \frac{2}{3}\alpha_{21}\left(2\alpha_{15} + \alpha_{19}\right)\left(\frac{\alpha_{18}^2\Lambda}{\Upsilon^-}\right)^2\frac{a\left(wb\right)^2}{f_2f_3c}H_2^\pm + \frac{4\alpha_{21}}{3}\left(\frac{\alpha_{18}\Lambda}{\Upsilon^-c}\right)^2\frac{wab}{f_3}H_3^\pm + \cr
&& - \frac{16\alpha_{21}\Upsilon^+}{9}\left(\frac{\Lambda}{\Upsilon^{-2}}\right)^2\left(\frac{wb^3}{f_3c^2}\right)^2H_4^\pm,
\end{eqnarray}\end{subequations}
\end{widetext}
where
\begin{widetext}
\begin{equation}
\Omega = \Lambda + \alpha_{21}, \quad \Upsilon^+ = \Lambda^2 +
4\alpha_{18}^2, \quad \Upsilon^- = \alpha_{21}\Lambda - 4\alpha_{18}^2.
\end{equation}
\end{widetext}
{\footnotesize
\begin{center}
\begin{table}[hbt]
\caption{\label{tab1}\footnotesize\baselineskip = 12pt Neutral standard gauge boson couplings with $H_2^{\pm\pm}$ and $H_3^{\pm\pm}$ in minimal 3-3-1 model from the covariant derivatives (\ref{cov}). It is necessary include the factor $ie\left(p - q\right)_\mu$, where $p_\mu$ and $q_\mu$ are four-moments.}
\begin{ruledtabular}
\begin{tabular}{c|ccc}
& $H_2^{\pm\pm}H_2^{\mp\mp}$ & $H_2^{\pm\pm}H_3^{\mp\mp}$ &
$H_3^{\pm\pm}H_3^{\mp\mp}$ \\
\hline
$\gamma$ & $\left(4\Lambda bc/\alpha_{21}\right)^2$ & $8\sqrt{2}\Lambda bc/\alpha_{21}$ & $2$ \\
$Z_\mu$ & $-1/2s_Wc_W$ & $2\sqrt{2}\Lambda\left(1 - 4s_W^2\right)bc/\alpha_{21}s_Wc_W$ & $-2s_W/c_W$
\end{tabular}\end{ruledtabular}\end{table}\end{center}}
Thus, Tables \ref{tab1} and \ref{tab2} allow us to compare the coupling strengths given by the 3-3-1 model with those coming from HTM (Table \ref{tab3}) relevant to the diagrams of Fig. 1. Therefore, we can see that $\gamma H_3^{++}H_3^{--}$, $Z_\mu H_2^{++}H_2^{--}$, $Z_\mu H_3^{++}H_3^{--}$ and $W_\mu H_2^{--}H_1^+$ interactions have practically the same strength that their correspondent HTM couplings. The other interactions depend of 3-3-1 parameters and so, the comparison with HTM predictions is less objective. In fact they are suppressed by powers of $1/w$, but in principle these suppression factors can be counterbalanced by the other Higgs sector parameters. \par
Signs of DCHBs of the 3-3-1 model would be affected by noise from the SM of the same way as the other models. These effects were studied in several papers and have been consistently shown that they are insignificant. Yet, often a small contamination is admitted. It stems from the confusion of jets with leptons from the decay products and misidentification of the electric charge. Events in which two vector bosons decay into leptons may also contribute. For the event selection criteria and methods of suppression of these backgrounds see, for example, Refs. \cite{HM07,CM11b}. The 3-3-1 model itself has several scalars and vector fields which decay into charged leptons and that, in principle, could mask the signal from DCHB if their masses are close to the Fermi scale. However, the typical model couplings are suppressed by powers of $1/w$, and most of them involve exotic fermions. The most dangerous case would be $U^{\pm\pm} \to \ell^\pm\ell^\pm$, where $U^{\pm\pm}$ are gauge bosons, because of its resemblance to the decays of $H_ {2,3}^{\pm\pm}$. However, this signal can be eliminated through its angular distribution, provided that the scalars decay isotropically.
\begin{widetext}
{\footnotesize
\begin{center}
\begin{table}[hbt]
\caption{\label{tab2}\footnotesize\baselineskip = 12pt Coupling as those
of the Table I, but for the standard charged gauge boson coupling with doubly and singly charged Higgs.}
\begin{ruledtabular}
\begin{tabular}{c|cccc}
& $H_1^\pm$ & $H_2^\pm$ & $H_3^\pm$ & $H_4^\pm$ \\
\hline
$W_\mu^\pm H_2^{\mp\mp}$ & $1/s_W$ &
$4\alpha_{18}\Lambda^2abc/\alpha_{21}\Upsilon^+s_W$ & $-
4\alpha_{18}\Lambda^2abc/\alpha_{21}\Upsilon^+s_W$ &
$16\alpha_{18}^3\Lambda^2b^5w/3\Upsilon^-s_Wcf_3$ \\
$W_\mu^\pm H_3^{\mp\mp}$ & $-2\sqrt{2}\Lambda bc/\alpha_{21}s_W$ &
$\sqrt{2}\alpha_{18}\Lambda a/\Upsilon^+s_W$ & $-
\sqrt{2}\alpha_{18}\Lambda a/\Upsilon^+s_W$ &
$4\alpha_{18}^3\alpha_{21}\Lambda b^4w/3{\Upsilon^-}^2s_Wc^2$
\end{tabular}\end{ruledtabular}\end{table}\end{center}}
\end{widetext}
This shows that the minimal 3-3-1 model is very competitive compared with HTM concerning DCHB searchings in the LHC under cross section level analysis. In a slightly different context, it was shown that the $H^{\pm\pm}_3$ can generates $\left(0.1 - 10\right)$ events/year ($\sqrt{s} = 7$ TeV) and $\left(100 - 1000\right)$ events/year ($\sqrt{s} = 14$ TeV) at the LHC \cite{RB11}. With regards the signal types, in the HTM the $BR\left(H^{\pm\pm} \to \ell^\pm{\ell^\prime}^\pm\right)$ can reach $\approx 100 \%$ for any charged lepton pairs for small triplet VEV values \cite{HM07}. On the other hand, the 3-3-1 model favors DCHBs decays into taus and does not allow decays of $H^{\pm\pm}$ into $\ell^\pm\ell^{\prime\pm}$ $\left(\ell \neq \ell^\prime\right)$, as discussed. The model predicts $BR\left[H_{2,3}^{\pm\pm} \to e^\pm e^\pm \left(\mu^\pm\mu^\pm\right) \left(\tau^\pm\tau^\pm\right)\right] \approx 8.21 \times 10^{-6} \% \left(0.35 \%\right) \left(99.65 \%\right)$
assuming the numerical values of the previous section. These results almost don't change if other parameter values are chosen in a realistic range. Therefore, a 3-3-1 DCHB with mass around $300$ GeV can gives events in a sufficient number to be discovered at the LHC thought decays into $\mu^\pm\mu^\pm$  pairs in three years of run \cite{AT11b}. Regarding the decays into $\tau^\pm\tau^\pm$, it include hadrons and neutrinos in their final products, which introduce additional difficulties. However, despite this drawback, the ATLAS and CMS Collaborations have developed high performance techniques for identification and reconstruction of $\tau$ pairs which can make the experiment possible \cite{CM11c}. \par
It is instructive to see why the HTM and minimal 3-3-1 model, both containing scalar fields in the triplet representation [SU(2) in the former case and SU(3) in the latter], make predictions with different accuracy levels. In HTM, the charged lepton and neutrino masses come from of independent scalar multiplets. In this case, there is a connection between the neutrino mixing matrix and the DCHB decay widths through Yukawa constants. This allows imposing constraints on neutrino masses through DCHB leptonic decay width measurements \cite{GS08, KR08, HM07, KO03}. On the other hand, the minimal 3-3-1 model also contains a scalar triplet that gives a tree-level mass to neutrinos when its neutral component develops a VEV.{\footnotesize
\begin{center}
\begin{table}[hbt]
\caption{\label{tab3}\footnotesize\baselineskip = 12pt HTM couplings relevant for DCHB searches at the LHC \cite{AB11}. $v_T$ is the HTM triplet VEV. A factor of $ie\left(p - q\right)_mu$ was omitted.}
\begin{ruledtabular}
\begin{tabular}{ccc}
$\gamma H^{\pm\pm}H^{\mp\mp}$ & $Z_\mu H^{\pm\pm}H^{\mp\mp}$ & $W_\mu^\pm H^{\mp\mp}H^\pm$ \\
\hline
$2$ & $\left(1 - 2s_W^2\right)/s_Wc_W$ & $-1/s_W\sqrt{1 + 2v_T^2/v_W^2}$
\end{tabular}\end{ruledtabular}\end{table}\end{center}} However, it is embedded in the sextet of scalar fields (\ref{S}), which breaks down as ${\bm 6} = {\bm 3} \oplus {\bm 2} \oplus {\bm 1}$. In the original minimal 3-3-1 model, the representation ${\bm 2}$ is responsible for generating the electron mass and for breaking the degenerescence between the muon and tau masses. This complements the charged leptonic mass spectrum generated by the triplet $\eta$ in Eqs. (\ref{tri}) (see details in Ref. \cite{FH93}). However, the lepton masses can be completely generated only by the sextet (\ref{S}). Hence, we eliminate the leptonic coupling with scalar triplet $\eta$ to obtain the Yukawa Lagrangian (\ref{L}). Thus, when we diagonalize the charged lepton mass matrix, all Yukawa interactions become diagonal. This does not happen in HTM because of the independence between the scalar multiplets which generate the lepton masses. \par
{\it Conclusions.}$-$ We discuss here some interesting properties of DCHBs of the minimal 3-3-1 model which are not seen in other models. In particular, it may be noted that these DCHBs decay almost exclusively in pairs $\tau^\pm\tau^\pm$ $\left[BR\left(H_{2,3}^{\pm\pm} \to \tau^\pm\tau^\pm\right) > 99.5 \% \right]$. The rate of decay in $W^\pm W^\pm$ is highly suppressed or absent, because the corresponding coupling strength is proportional to the VEV of the Higgs which gives mass to neutrinos. Moreover, due to a property of the gauge symmetry of the model, the Yukawa couplings written in terms of physical fields are diagonal, which leads to simple relations between the masses, VEVs and decay widths. Unfortunately, because of the high preference for decays into $\tau^\pm\tau^\pm$, these widths are difficult to be measured, but apparently not impossible. In fact, decays into $\mu^\pm\mu^\pm$ could be identified at the LHC for DCHB masses of the order of 300 GeV in about three years of operation. Decays into $e^\pm e^\pm $ are very rare $\left[BR\left(H_{2,3}^{\pm\pm} \to e^\pm e^\pm\right) \sim 10^{-5}\right]$. The channel $\tau^\pm\tau^\pm$ is difficult to be identified because the $\tau$ lepton decay into neutrinos and hadrons. However, there is much interest in studies of the $\tau$ decays and, as discussed throughout this Letter, the ATLAS and CMS Collaborations are making efforts to put into practice high performance techniques for identification and reconstruction of the $\tau$ signals \cite{CM11c}. \par
It is interesting to notice that the HTM has fewer parameters when compared to the minimal 3-3-1. However, it has been established here that in the case of the DCHB decays into charged leptons, the minimal 3-3-1 model makes predictions more clearly defined. Therefore, the prediction ability of a model is not always directly related to the amount of fields and/or free parameters, as commonly thought. In fact, what determines the model power prediction are the fundamental symmetries and the representation content selected, which governs the coupling structures. As another example, the 3-3-1 model Ref. \cite{PT93} has the same fundamental symmetry but different leptonic and Higgs content and fewer parameters if compared with the model of Ref. \cite{FH93, FR92}. However, through the DCHB decays into charged leptons, it gives more uncertain results.\par
The features responsible to these peculiarities of the minimal 3-3-1 model are that (a) lepton and anti-lepton pairs are include in the same leptonic representation content in the SU(3) triplet (\ref{psi}) and (b) all the leptons get their masses from the Higgs fields in the same SU(3) representation expressed in the sextet (\ref{S}). This leads to a relative {\it economy} of Yukawa constants, so that the charged lepton
mass matrix and the interactions of leptons with the scalars are simultaneously diagonalized.

\acknowledgments

The author would like to thank E. C. F. S. Fortes for reading the manuscript and the Instituto de F\'\i sica Te\'orica of UNESP for hospitality. This work was supported in part the Brazilian agencies CNPq (research fellowship) and FAPESP (financial support for the research, Processo No. 2009/02272-2).

\end{document}